\documentclass[10pt]{article}
\usepackage{epsfig}

\begin{document}


   \title{\LARGE \bf  Non-Singular Stationary Global Strings\\}

   \author{Y. Verbin$^a$\thanks{Electronic address: verbin@oumail.openu.ac.il}
  \large
  \setcounter{footnote}{3}
 and A.L. Larsen$^b$\thanks{Electronic address: all@fysik.sdu.dk}
 }
 \date{ }
   \maketitle
   \centerline{$^a$ \em Department of Natural Sciences, The Open
University of Israel}  \centerline{\em P.O.B. 39328, Tel Aviv
61392, Israel}
 \vskip 0.4cm
\centerline{$^b$ \em Department of Physics, University of Southern
Denmark, }
   \centerline{\em Campusvej 55, 5230 Odense M, Denmark }

       \vskip 1.1cm

   \begin{abstract}
A field-theoretical model for non-singular global cosmic strings
is presented. The model is a non-linear sigma model with a
potential term for a self-gravitating complex scalar field.
Non-singular stationary solutions with angular momentum and
possibly linear momentum are obtained by assuming an oscillatory
dependence of the scalar field on $t$, $\varphi$ and $z$. This
dependence has an effect similar to gauging the global $U(1)$
symmetry of the model, which is actually a Kaluza-Klein reduction
from four to three spacetime dimensions. The method of analysis
can be regarded as an extension of the gravito-electromagnetism
formalism beyond the weak field limit. Some $D=3$ self-dual
solutions are also discussed. \\
   \end{abstract}


\section{Introduction}\label{Introduction}
\setcounter{equation}{0}

Cosmic strings are a special kind of topological defects
\cite{VilSh}, which could have been formed in a symmetry breaking
phase transition in the early universe. Often, the broken symmetry
is assumed to be local, but the possibility of global strings has
also been extensively discussed \cite{VilSh}. The prototype is
Cohen and Kaplan's solution
\cite{CohenKaplan,Gregory1988,GibbonsEtAl1989} of a complex scalar
with a $U(1)$ symmetry breaking potential. The string solution
turns out to have a repulsive gravitational field with a curvature
singularity at a finite distance from the axis. The singularity is
unavoidable as was shown by Gibbons et al. \cite{GibbonsEtAl1989}
for the standard Lagrangian of a complex field with a non-negative
potential term, under the assumptions that the solutions are
static and have symmetry under boosts along the string axis. This
result may be regarded as the analog in the gravitational case of
the $D=2+1$ dimensional version of Derrick's theorem
\cite{Rajaraman}, excluding the existence of static global strings
with finite energy per unit length in flat space.

It was shown explicitly that the singularity of the Cohen-Kaplan
solution can be avoided if time dependence is allowed, which
results a string solution with de-Sitter like expansion along the
string axis \cite{Gregory1996}. An attempt to get a static
non-singular global string was tried \cite{SenBanerjee} by
relaxing the symmetry under boosts parallel to the string axis
($z$), but without relaxing the corresponding condition
$(T_t^t=T_z^z)$ in the string core, so the solution found there has
lost its interpretation as a global string gravitational field.

There are additional ways to overcome the ``no-go'' theorems and
find string solutions. All of them involve modifications of the
Cohen-Kaplan model. Probably the most well-known is gauging the
$U(1)$ symmetry group to become local, thus obtaining the Abelian
Higgs model. The main consequence of this modification is that the
long range massless Goldstone field transmutes into the
longitudinal component of a short range massive vector field. This
model admits string solutions (flux tubes) in flat space and in
the self-gravitating case. However, we will not discuss this
possibility here since almost all that can be said about it, has
been already said \cite{VilSh}. Other ways are either restricted
to the flat space case, or to the gravitating case or are valid in
both. One is the Q-ball solution \cite{Coleman85}, which overcomes
Derrick's theorem by allowing oscillatory time dependence of the
complex scalar, so as to avoid the staticity assumption without
introducing time dependence into the energy-momentum tensor and
other observables. In addition, the potential term is modified in
such a way that it has another symmetric local minimum. From a
more physical point of view, these modifications result a
non-vanishing value of the conserved global $U(1)$ charge, which
stabilizes the spherical solutions against collapse. The
string-like analogue (``Q-string'') is similarly stabilized by a
finite global $U(1)$ charge per unit length (see however
\cite{MacKPar2001} for a different situation). Spinning versions
have been also recently considered \cite{VolWohn}. Adding gravity
into the Q-ball system does not change the overall picture of the
solutions, but rather yields the so-called Q-stars \cite{Lynn89}.

A second way which is also based on the existence of a global
$U(1)$ charge but in which gravity is indispensable, gives the
boson stars \cite{Liddle1992,Jetzer1992,LeePang1992}.

It is also possible to get similar ``non-topological solitons'' in
systems with more than one complex scalar field
\cite{LeePang1992}, but since we want to stick to a minimal field
content in our model, we will not discuss those as well.

The simplest way to add a topological charge is to assume for the
scalar field a ``hedgehog configuration'', where it tends
asymptotically to symmetry breaking minima of the potential,
mapping this way spatial infinity into the vacuum manifold. In the
original $U(1)$ system, static string-like solutions fall indeed
into homotopy classes characterized by an integer winding number,
but such solutions are excluded by the ``no go theorems''
mentioned above. However, we may think of a second version of a
``hedgehog configuration'', where we turn the target space from a
trivial vector space into a compact manifold of a non-linear sigma
model. Since we are after string-like solutions, the scalar field
defines a mapping to target space from the compactified
$\mbox{\textrm{R}}^{2}$, which is transverse with respect to the
string axis. If we take the simplest target space
$\mbox{\textrm{S}}^{2}$, the kinetic part of the Lagrangian will
be the usual one of the $O(3)$ non-linear sigma model.  The
potential term will necessarily break explicitly the $O(3)$
symmetry, but we will assume it is still $U(1)$-symmetric. It will
be also taken until further notice to be non-negative with minima
at $U=0$. In this form of hedgehog configuration,  the scalar field
should tend asymptotically to a $U(1)$-symmetric minimum in order
to correspond to a linear defect, which will be both localized
enough and have a topological charge.

To make the discussion more concrete, we write down the action in
the general form
\begin{equation}
S=\int d^Dx\sqrt{|g|}\left(\frac{1}{2}{\cal
E}(|\Phi|)\nabla_\mu\Phi^* \nabla^\mu\Phi -
U(|\Phi|)+\frac{1}{2\kappa}R\right) \label{action1}\end{equation}
where $\kappa$ is the $D$ dimensional gravitational constant (we
will discuss only $D=3,4$) and ${\cal E}(|\Phi|)$ is a
non-negative dimensionless function, which serves as a Weyl factor
of a conformally-flat target space metric of the non-linear sigma
model. The $\mbox{\textrm{S}}^{2}$ target space is obtained by
\begin{equation}
{\cal E}(|\Phi|)=\frac{1}{(1+|\Phi|^2/\mu^2)^2}
\label{SphereConfFactor}
\end{equation}
where $\mu$ is the diameter of the 2-sphere and sets the energy
scale. In $D$ dimensions it has dimensions of energy to the power
$(D-2)/2$. More conventions: $\nabla_\mu$ is a covariant
derivative, signature $(+,-,\dots,-)$ and the Riemann tensor is
$R^\kappa_{\lambda\mu\nu}=
\partial_\nu\Gamma^\kappa_{\lambda\mu}-\partial_\mu\Gamma^\kappa_{\lambda\nu}+\dots$.

Note that the ``linear'' Higgs (or rather Goldstone --- but we
will use the former term) system is recovered either by taking
${\cal E}(|\Phi|)=1$, or using the sigma model Lagrangian in the
limit of small fields or formally for $\mu\rightarrow\infty$.

In flat spacetime this model has well-known static string-like
solutions, but only with a vanishing potential -- again a simple
consequence of Derrick's theorem. This is the original $O(3)$
sigma model \cite{Rajaraman}. We will still use the term ``sigma
model'' for the action (\ref{action1}), but refrain from
characterizing it by the symmetry group, since it is (explicitly)
broken by the potential term to $U(1)$.

The flat space string-like solutions were found to be unstable and
may either shrink or spread out
\cite{LeeseEtAl1990,PietteZakr1990}. They may be however
stabilized by allowing (as above) an oscillatory time dependence
and adding a potential term. This produces the so-called
``Q-lumps'', which are stable rotating topological string-like
solutions \cite{Leese1991,Ward2003}. As described, these solutions
have oscillatory time as well as angular dependence (around the
string axis): $\Phi = \mu f(r)e^{i(m\varphi-\omega t)}$, where $m$
is an integer and $\omega$ real. The non-vanishing global $U(1)$
charge results from the oscillatory time dependence, while it is
the mixed angular and time dependence which produces the
non-vanishing angular momentum density, $T^t_\varphi$. The
topological winding number is $m$.

The $O(3)$ sigma model may be also stabilized by a gravitational
field in the form of a cosmological expansion \cite{Ward2002}.
Closer to our present purpose is the Comtet-Gibbons solution
\cite{ComtetGibbons1988}, which is a static cylindrically symmetric
(self-dual) solution of a self-gravitating $O(3)$ sigma model
(with a vanishing potential).

In this paper, we fill some of the gaps which are evident from the
above discussion, and find new non-singular global string
solutions, which are topologically non trivial: self-gravitating
Q-lumps and generalizations which introduce also a momentum
current along the string axis - a possibility which has not been
considered before. This is reflected by $T^t_\varphi \neq 0$ in
case of rotation or $T^t_z \neq 0$ in case of translational motion
along the string axis. This will be done by taking the more
general oscillatory behavior $\Phi = \mu f(r)e^{i(qz-\omega t
+m\varphi)}$. Some non-topological solutions like self-gravitating
spinning Q-strings will be discussed in a future publication.

The expectation that one may get non-singular stationary
gravitational fields in this way, is based on the observation that
introducing an oscillatory $z$ and/or $t$-dependence is consistent
with the existence of two commuting Killing vectors, which generate
translations in time and along the string axis, and has the effect
of gauging the global symmetry and turning the global sigma model
into a gauged one. This way, the original $D=4$ gravitating global
sigma model is reduced in a Kaluza-Klein way into a $D=3$ gauged
sigma model coupled with $D=3$ gravity and electromagnetism (and a
dilaton). Since it is known that the gravitating Nielsen-Olesen
flux tube produces a non-singular gravitational field, even when a
dilaton coupling is allowed
\cite{GregorySantos1997,Santos2001,VMLC}, there is a good chance
it will share this characteristic with the type of solutions
described above. As the dilaton introduced by the Kaluza-Klein
reduction is also a long range field, the gravitational field of
the global string solutions obtained in our model will be finite
but not asymptotically flat. Only the pure $D=2+1$ system
discussed in sec. \ref{Three-Dimensional System} will admit
asymptotically flat global vortices.

We proceed and end this section by writing the field equations of
our model. They are the following generalization of Klein-Gordon
equation
\begin{equation}
{\cal E}(|\Phi|)\nabla_\mu\nabla^\mu\Phi +
\frac{\Phi^*}{2|\Phi|}\frac{d{\cal E}} {d|\Phi|}\nabla_\mu\Phi
\nabla^\mu\Phi+ \frac{\Phi}{|\Phi|}\frac{dU}{d|\Phi|}=0
\label{KGEqD}\end{equation} together with the Einstein equations
(in term of the Einstein tensor $G_{\mu\nu}$)
\begin{equation}
-\frac{1}{\kappa}G_{\mu\nu}=T_{\mu\nu}= \frac{1}{2}{\cal
E}(|\Phi|)\left[\partial_\mu\Phi^*\partial_\nu\Phi+
\partial_\nu\Phi^*\partial_\mu\Phi -(\nabla_\lambda\Phi^*)(\nabla^\lambda\Phi)
g_{\mu\nu}\right]+U(|\Phi|)g_{\mu\nu}
\label{Einstein}\end{equation} which are equivalent to
\begin{equation}
\frac{1}{\kappa}R_{\mu\nu}=\frac{2}{D-2}U(|\Phi|)g_{\mu\nu}-
\frac{1}{2}{\cal E}(|\Phi|)
\left[\partial_\mu\Phi^*\partial_\nu\Phi+
\partial_\nu\Phi^*\partial_\mu\Phi \right]
\label{EinsteinSigma}\end{equation}

\section{Field Equations for Sigma Model Global Strings}
\setcounter{equation}{0} \label{Sigma Model Strings}

We start with the $D$-dimensional action (\ref{action1}) and look
for string-like solutions, where the metric tensor is written in
the form
\begin{eqnarray}
ds^2=g_{\mu\nu}(x^\lambda)dx^\mu
dx^\nu=e^{2\sigma(x^\gamma)}h_{\alpha\beta}(x^\gamma)dx^\alpha
dx^\beta
-e^{-2(D-3)\sigma(x^\gamma)}[dz+A_{\alpha}(x^\gamma)dx^\alpha]^2.
\label{lineelGen1}\end{eqnarray} The indices $\alpha,\beta...$ go
from 0 to $D-2$, and we denote $x^{D-1} = z$. This is just the
Kaluza-Klein reduction from $D$ to $D-1$ dimensions. If we also
take $\Phi=\psi(x^\alpha) e^{iqz}$, the $D$-dimensional gravitating
sigma model with a global U(1) symmetry turns into a gravitating
gauged sigma-Maxwell-dilaton system in $D-1$ dimensions, as it is
evident from the Lagrangian which becomes:
\begin{eqnarray}
\sqrt{|g|}\left(\frac{1}{2}{\cal E}(|\Phi|)\nabla_\mu\Phi^*
\nabla^\mu\Phi - U(|\Phi|) + \frac{1}{2\kappa}R(g)\right)=\nonumber \\
\sqrt{|h|}\left(\frac{1}{2}{\cal E}(|\Phi|)[D_\alpha\psi^*
D^\alpha\psi - q^2e^{2(D-2)\sigma}|\psi|^2]-e^{2\sigma}U(|\psi|)\right)+\nonumber\\
\frac{\sqrt{|h|}}{2\kappa}\left(R(h)+(D-2)(D-3)\nabla_\alpha\sigma
\nabla^\alpha\sigma-
\frac{1}{4}e^{-2(D-2)\sigma}F_{\alpha\beta}F^{\alpha\beta}+
2(D-2)\nabla_\alpha\nabla^\alpha\sigma\right)
\label{RedLag}\end{eqnarray}
after we define a usual covariant derivative
$D_\alpha =\partial_\alpha-iqA_\alpha$ and field strength
$F_{\alpha\beta}=\partial_\alpha A_\beta-\partial_\beta A_\alpha$. The last term
can obviously be omitted. All geometric quantities are related to the $(D-1)$ metric
$h_{\alpha\beta}$.

In the next step we assume stationarity, i.e. parametrize the scalar field and
the $D-1$ metric as:
\begin{eqnarray}
\psi=\phi(x^k)e^{-i\omega t} ,\ \ \   h_{\alpha\beta}(x^\gamma)dx^\alpha dx^\beta =
N^2(x^k)[dt+L_i(x^k)dx^i]^2-\gamma_{ij}(x^k)dx^idx^j
\label{lineelD-1}\end{eqnarray}
where the indices $i,j...$ go from 1 to $D-2$. Now we do not write the
Lagrangian or the action in the $D-2$ dimensional notation, but assume $D=4$
and write directly
the field equations in terms of the following:
\begin{itemize}
\item $\nabla_i$  the covariant derivative with respect to the
2-dimensional metric $\gamma_{ij}$,
\item $R_{ij}(\gamma)$
the corresponding Ricci tensor,
\item
$L_{ij}=\partial_iL_j-\partial_jL_i=\ell\sqrt{|\gamma|}\epsilon_{ij}$.
\item $F_{0i}=NE_{i}, \ \ \ F^{0i}=-D^{i}/N$
\item $F_{ij}=-B\sqrt{|\gamma|}\epsilon_{ij}, \ \ \
F^{ij}=-H\epsilon^{ij}/\sqrt{|\gamma|}$
\item $\bar{D}_j =\partial_j-i[qA_j-(\omega+qA_0)L_j]=
\partial_j-iq\bar{A}_j$
\item $\bar{J}_k = -\frac{i}{2} {\cal E}(|\phi|)(\phi^{*}\bar{D}_j \phi-
\phi\bar{D}_j \phi^{*})$
\end{itemize}
We use the usual convention of raising and lowering indices by
$\gamma_{ij}$ and its inverse. Note the distinction between
$D^{i}$ which is an ``electric'' (displacement) field (obeying
$D_{i}=E_{i}+NH\sqrt{|\gamma|}\epsilon_{ij}L^{j}$), and
$\bar{D}^i$ which is a covariant derivative. We will also use
$\textbf{E}^2$ to denote $E_iE^i$. This way of dealing with the
field equations is essentially an extension of the
gravito-electromagnetism formalism beyond the weak field limit,
where it is usually used \cite{Mashhoon2003,TartRugg2003}. The
components of $F_{\alpha\beta}$ will be referred to as
gravito-electric and gravito-magnetic.

Einstein equations split therefore according to the components
(00), $(zz)$, $(i0)$, $(iz)$, $(0z)$ and $(ij)$. The first 5 are:
\begin{eqnarray}
\frac{1}{N}\nabla_i \nabla^i N+\frac{N^2\ell^2}{2}-
\frac{H^2}{2}e^{-4\sigma}&=&-\kappa\left(2e^{2\sigma}U(|\phi|)+
\left(q^2e^{4\sigma}-\frac{(\omega+qA_0)^2}{N^2}\right)|\phi|^2{\cal E}(|\phi|)\right)
\label{eqN}\\
\frac{1}{N}\nabla_i (N\nabla^i \sigma)+
\frac{H^2-\textbf{E}^2}{2}e^{-4\sigma}&=&\kappa\left(e^{2\sigma}U(|\phi|)+
q^2e^{4\sigma}|\phi|^2{\cal E}(|\phi|)\right)
\label{eqsig}\\
\frac{\epsilon^{ij}}{N\sqrt{|\gamma|}}\partial_j(N^3 \ell)-
\frac{e^{-4\sigma}HN\epsilon^{ij}}{\sqrt{|\gamma|}}E_j &=&
2\kappa(\omega+qA_0)\bar{J}^i
\label{eqell}\\
\frac{\epsilon^{ij}}{N\sqrt{|\gamma|}}\partial_j(NHe^{-4\sigma})&=&
-2\kappa q\bar{J}^i
\label{eqH}\\
\frac{1}{N}\nabla_i(e^{-4\sigma}D^i) &=& 2\kappa q
\left(L_{i}\bar{J}^i- \frac{\omega+qA_0}{N^2}|\phi|^2{\cal
E}(|\phi|)\right) \label{eqD}
\end{eqnarray}
and the $(ij)$ components are:
\begin{eqnarray}
R_{ij}(\gamma)+\frac{1}{N}\nabla_i\nabla_j N -\frac{N^2\ell^2}{2}\gamma_{ij}+
2\partial_i \sigma \partial_j \sigma+
\frac{1}{2}(\textbf{E}^2 \gamma_{ij}-E_i E_j)e^{-4\sigma}=\nonumber \\
-\kappa\left(\left(2e^{2\sigma}U(|\phi|)+
q^2e^{4\sigma}|\phi|^2{\cal E}(|\phi|)\right)\gamma_{ij}+
\frac{1}{2}{\cal E}(|\phi|)(\bar{D}_i\phi^*\bar{D}_j\phi+
\bar{D}_j\phi^*\bar{D}_i\phi)\right)
\label{eqRedEinst}\end{eqnarray}
The equation for the scalar field is
\begin{eqnarray}
\bar{D}_i\bar{D}^i\phi+
\frac{\phi^*}{2|\phi|{\cal E}}\frac{d{\cal E}}{d|\phi|}\bar{D}_i\phi\bar{D}^i\phi+
\left(\frac{(\omega+qA_0)^2}{N^2}-q^2e^{4\sigma}\right)
\left(1+\frac{|\phi|}{2{\cal E}}\frac{d{\cal E}}{d|\phi|}\right)\phi-
\frac{e^{2\sigma}}{|\phi|{\cal E}}\frac{dU}{d|\phi|}\phi=0
\label{eqRedKG}\end{eqnarray}

Another equation which is not independent, but will be useful is the $(ij)$ component
of (\ref{Einstein}):
\begin{eqnarray}
\frac{1}{N}(\nabla_i\nabla_j N-\gamma_{ij}\nabla_k\nabla^k N )
-\frac{N^2\ell^2}{4}\gamma_{ij}+
2\partial_i \sigma \partial_j \sigma -
\gamma_{ij} \gamma^{kl} \partial_k \sigma \partial_l \sigma +\nonumber\\
\frac{e^{-4\sigma}}{2}\left(\frac{H^2+\textbf{E}^2}{2} \gamma_{ij}-E_i E_j\right)
=\nonumber \\
-\frac{\kappa}{2}{\cal E}(|\phi|)(\bar{D}_i\phi^*\bar{D}_j\phi+
\bar{D}_j\phi^*\bar{D}_i\phi)+\nonumber\\
\frac{\kappa}{2}\left({\cal E}(|\phi|)
\bar{D}_k\phi^*\bar{D}^k\phi\ +2e^{2\sigma}U(|\phi|)+
\left(q^2e^{4\sigma}-\frac{(\omega+qA_0)^2}{N^2}\right)
|\phi|^2{\cal E}(|\phi|)\right)\gamma_{ij}
\label{eqGij}\end{eqnarray}

\section{Three-Dimensional System}
\setcounter{equation}{0} \label{Three-Dimensional System} In order
to get a clearer picture of the system, it is useful to deal with a
lower dimensional case if possible. Although this set of field
equations is genuinely four-dimensional, it contains also in a
special limit the field equations for the $D=3$ system. These are
equations (\ref{eqN}), (\ref{eqell}), (\ref{eqRedEinst}),
(\ref{eqRedKG}) with the substitutions: $q=0$, $\sigma=0$ and
$A_{\alpha}=0$, which imply $E_i=D_i=0$ and $B=H=0$. We get
therefore:
\begin{eqnarray}
\frac{1}{N}\nabla_i \nabla^i N+\frac{N^2\ell^2}{2}&=&-\kappa\left(2U(|\phi|)
-\frac{\omega^2}{N^2}|\phi|^2{\cal E}(|\phi|)\right)
\label{eqN3D}\\
\frac{\epsilon^{ij}}{N\sqrt{|\gamma|}}\partial_j(N^3 \ell) &=&
2\kappa\omega\bar{J}^i
\label{eqell3D}\\
R_{ij}(\gamma)+\frac{1}{N}\nabla_{i}\nabla_{j}N -\frac{N^2\ell^2}{2}\gamma_{ij}&=&
-\kappa\left(2U(|\phi|)\gamma_{ij}+
\frac{1}{2}{\cal E}(|\phi|)(\bar{D}_i\phi^*\bar{D}_j\phi+
\bar{D}_j\phi^*\bar{D}_i\phi)\right)
\label{Einst3D}\end{eqnarray}
and for the scalar field:
\begin{eqnarray}
\bar{D}_i\bar{D}^i\phi+
\frac{\phi^*}{2|\phi|{\cal E}}\frac{d{\cal E}}{d|\phi|}\bar{D}_i\phi\bar{D}^i\phi+
\frac{\omega^2}{N^2}\left(1+\frac{|\phi|}{2{\cal E}}\frac{d{\cal E}}
{d|\phi|}\right)\phi-
\frac{1}{|\phi|{\cal E}}\frac{dU}{d|\phi|}\phi=0
\label{eqKG3D}\end{eqnarray}

Note that a solution of the $D=3$ system is not automatically a solution to the $D=4$
system with the above-mentioned substitutions. The main reason is eq (\ref{eqsig})
which gives an additional condition, which is consistent with the $D=3$ theory only in the
absence of self-interaction (i.e. $U=0$). On the other hand, the fact that there is an
explicit dependence in the Einstein equations upon the dimensionality $D$, does not
hinder the possibility to use them for the $D=3$ field equations. Accordingly, the $D=3$
analogue of eq (\ref{eqGij}) is:
\begin{eqnarray}
\frac{1}{N}(\nabla_i\nabla_j N-\gamma_{ij}\nabla_k\nabla^k N )
-\frac{N^2\ell^2}{4}\gamma_{ij}=
-\frac{\kappa}{2}{\cal E}(|\phi|)(\bar{D}_i\phi^*\bar{D}_j\phi+
\bar{D}_j\phi^*\bar{D}_i\phi)+\nonumber\\
\frac{\kappa}{2}\left({\cal E}(|\phi|)
\bar{D}_k\phi^*\bar{D}^k\phi\ +2U(|\phi|)-\frac{\omega^2}{N^2}
|\phi|^2{\cal E}(|\phi|)\right)\gamma_{ij}
\label{eqGij3D}\end{eqnarray}

\section{Self-Dual Solutions}
\setcounter{equation}{0} \label{Self-Dual Solutions}

It is known that the field equations simplify considerably if self-duality is
possible (see e.g. ref. \cite{VML} and references therein).
So we will start in the simplest case of self-dual solutions in the $D=3$ system.
The self-dual solutions obey
\begin{equation}
N=1 , \ \ \ \bar{D}^i\phi =i\eta\frac{\epsilon^{ij}}{\sqrt{|\gamma|}}
\bar{D}_j\phi
 \label{SD3D}\end{equation}
 where $\eta=\pm 1$ corresponds to self-dual or anti self-dual solutions.
 In the following we will refer to both as ``self-dual''.
 This is consistent with eq. (\ref{eqN3D}) or (\ref{Einst3D}), provided the
 following relation holds:
\begin{equation}
\frac{\ell^2}{2\kappa}+2U(|\phi|)-\omega^2|\phi|^2{\cal E}(|\phi|)=0.
 \label{SDconsistency}\end{equation}
This relation is also consistent with eq (\ref{eqKG3D}).
Since the current also simplifies considerably for self-dual solutions, we get from
eq (\ref{eqell3D})
\begin{equation}
\partial_i \ell = 2\eta\kappa\omega |\phi|{\cal E}(|\phi|)\partial_i |\phi|
 \label{SDell3D}\end{equation}
This equation can be integrated to give $\ell(|\phi|)$ for any
given ${\cal E}(|\phi|)$. In the most familiar case of the
$\mbox{\textrm{S}}^2$ sigma model defined by eq.
(\ref{SphereConfFactor}), the integration is trivial and we find
the following form of potential term, which will give rise to
localized self-dual solutions:
\begin{equation}
U_{SD}(|\phi|)=\frac{\omega^2\mu^2(|\Phi|^2/\mu^2-\kappa\mu^2/2)}
{2(1+|\Phi|^2/\mu^2)^2}
\label{SDPot}\end{equation}
We chose the integration constant such that $\ell$ will vanish
asymptotically, i.e. for fields that satisfy $|\Phi|\rightarrow\infty$ as
$r\rightarrow\infty$. It is interesting to note that the linear sigma model
(${\cal E}(|\phi|)=1$) needs for self-duality a potential which is unbounded from
below, which is a property shared by other systems \cite{VML}.

Since target space is a sphere, it is natural and useful to use an
``angular'' field defined by $|\Phi|=\mu \tan(\Theta/2)$. In
terms of this field, we get for example ${\cal E}=\cos^4(\Theta/2)$
and see that $|\Phi|\rightarrow\infty$ is actually just the
``south pole'' of target space, $\Theta=\pi$. Thus, the self-duality potential can
be rewritten as:
\begin{equation}
U_{SD}(\Theta)= \frac{\omega^2\mu^2}{8}\left( \sin^2 \Theta -
2\kappa\mu^2 \cos^4 (\Theta/2)\right )
\label{SDPotTheta}\end{equation}

This generalizes the flat space self-dual Q-lumps \cite{Leese1991,Ward2003}
to be self-gravitating. The flat space limit is reproduced by taking $\kappa=0$.
 From a different direction, this can be regarded as a
spinning generalization of the self-dual Comtet-Gibbons solutions
\cite{ComtetGibbons1988}. Those are reproduced for $\omega=0$.
Explicit solutions are presented in section \ref{SpecSol}.

\section{Rotational Symmetry}
\setcounter{equation}{0} \label{Rotational Symmetry}

We now return to $D=4$  and specialize to rotational symmetry. We
are left with the 2-metric
\begin{equation}
\gamma_{ij}dx^idx^j= \alpha^2(r)dr^2+\beta^2(r)d\varphi^2
 \label{lineelRotSym}\end{equation}
where we postpone the gauge fixing to a later stage. The vector
$A_{\alpha}$ will have only two non-vanishing components
$A_{0}(r)$ and $A_{\varphi}(r)$, from which we derive the
gravito-electric and gravito-magnetic fields $E_{r}(r)$ and
$B(r)$. Similarly, $L_i$ will have only one non-vanishing
component namely $L_{\varphi}(r)$; thus $\ell$ will also depend on
$r$ only. We will also use $\bar{A}_\varphi$ defined by
$q\bar{A}_\varphi=qA_\varphi-(\omega+qA_0)L_{\varphi}$.

We define a dimensionless potential function $u$ by
$U=\lambda\mu^4u(\Theta)$ where $\lambda$ is a dimensionless
coupling constant (in $D=4$), and  a dimensionless Newton's
constant $\gamma=\kappa\mu^2$ and get therefore the following set
of field equations for the $\mbox{\textrm{S}}^2$ model. From
equations (\ref{eqN})-(\ref{eqD}) we get
\begin{eqnarray}
\frac{1}{N\alpha\beta}\left(\frac{\beta N'}{\alpha}\right)'+
\frac{N^2\ell^2}{2}-\frac{H^2}{2}e^{-4\sigma}&=
&\gamma\left[\left(\frac{(\omega+qA_0)^2}{N^2}-q^2e^{4\sigma}\right)
\frac{\sin^2\Theta}{4}-2\lambda\mu^2e^{2\sigma}u(\Theta)\right]
\label{EinstN}\\
\frac{1}{N\alpha\beta}\left(\frac{\beta N\sigma'}{\alpha}\right)'
+\frac{H^2-(E_{r}/\alpha)^2}{2}e^{-4\sigma}&=&
\gamma\left[q^2e^{4\sigma}\frac{\sin^2\Theta}{4}+\lambda\mu^2e^{2\sigma}u(\Theta)\right]
\label{EinstSigm}\\
\frac{\beta}{N\alpha}\left(N^3\ell\right)' - \frac{\beta
N}{\alpha}HE_{r}e^{-4\sigma}&=& -\frac{\gamma}{2}
(\omega+qA_0)(m-q\bar{A}_\varphi)\sin^2\Theta\
\label{EinstL}\\
\frac{\beta}{N\alpha}\left(NHe^{-4\sigma}\right)' &=&
\frac{\gamma}{2} q(m-q\bar{A}_\varphi)\sin^2\Theta\
\label{EinstH}\\
\frac{1}{N\alpha\beta}\left(\frac{\beta}{\alpha}E_{r}e^{-4\sigma}\right)'
+ H\ell e^{-4\sigma}&=& -\frac{\gamma}{2N^2}
q(\omega+qA_0)\sin^2\Theta
 \label{EinsteinE}\end{eqnarray}
where the last one is a combination of (\ref{eqH}) and
(\ref{eqD}). From eq (\ref{eqRedEinst}) we take the equation for
$R(\gamma)$
\begin{eqnarray}
\frac{2}{\alpha\beta}\left(\frac{\beta'}{\alpha}\right)'-\frac{3N^2\ell^2}{2}
 +\frac{H^2+E_{r}^2/\alpha^2}{2}e^{-4\sigma}+2\frac{\sigma'^2}{\alpha^2}=\nonumber\\
-\gamma\left(\frac{\Theta'^2}{4\alpha^2}+
\left(\frac{(m-q\bar{A}_\varphi)^2}{\beta^2}+
\frac{(\omega+qA_0)^2}{N^2}+q^2e^{4\sigma}\right)\frac{\sin^2\Theta}{4}
+ 2\lambda \mu^2 e^{2\sigma} u(\Theta) \right)
 \label{EinsteinRotR}\end{eqnarray}
and the $(\varphi\varphi)$ equation:
 \begin{eqnarray}
\frac{1}{\alpha\beta N}\left(\frac{N\beta'}{\alpha}\right)'-\frac{N^2\ell^2}{2}
 +\frac{E_{r}^2}{2\alpha^2}e^{-4\sigma}=
-\gamma\left( \left(\frac{(m-q\bar{A}_\varphi)^2}{\beta^2}+q^2e^{4\sigma}\right)
\frac{\sin^2\Theta}{4} + 2\lambda \mu^2 e^{2\sigma}u(\Theta) \right)
 \label{EinsteinRotphiphi}\end{eqnarray}
For the scalar field we get:
\begin{equation}
\frac{1}{N\alpha\beta}\left(\frac{N\beta\Theta'}{\alpha}\right)'+
\left(\frac{(\omega+qA_0)^2}{N^2}-
\frac{(m-q\bar{A}_\varphi)^2}{\beta^2}-q^2e^{4\sigma}\right)
\frac{\sin(2\Theta)}{2}-4\lambda\mu^2 e^{2\sigma}\frac{du}{d\Theta}=0,
 \label{KGSpec}\end{equation}
and we have also to add the following relations:
 \begin{eqnarray}
E_{r}=-\frac{A_{0}'}{N} , \ \ \ H=\frac{1}{\alpha\beta }
(L_{\varphi}A_{0}'-A_{\varphi}') , \ \ \
\ell=\frac{L_{\varphi}'}{\alpha\beta}.
 \label{EHBell}\end{eqnarray}
For completeness we also note that $B=-A_{\varphi}'/\alpha\beta$.

Note that once we make a gauge choice, one of these equations
becomes dependent on the others. When we solve them numerically we
use eq. (\ref{EinsteinRotphiphi}) as a check.

Few soliton-like solutions to this system are already known in the
gravitating case -- all of them static $(\omega=q=0,
L_{\varphi}=A_{\varphi}=A_{0}=0)$. For the ``linear'' case (${\cal
E}(|\Phi|)=1$) with a  potential with a $U(1)$ symmetry breaking
minimum, there exists Cohen and Kaplan's global string
\cite{CohenKaplan} which is singular as already mentioned in the
introduction. In the non-linear sector, the Comtet-Gibbons
string-like solution \cite{ComtetGibbons1988} is a self-dual
solution with vanishing potential.

Remotely related solutions are the dilatonic-Melvin solution
\cite{DowkerEtAl1994,Santos2001}, and the three-dimensional
dilaton-Maxwell-Einstein system that were discussed by some
authors
\cite{Koikawa1997,Fernando1999,Fernando2002A,Fernando2002B}. The
main difference between our model and the above mentioned
dilaton-Maxwell-Einstein papers, is that none of them has suggested
a possible source for the angular momentum or electromagnetic
fields of those solutions. It may be thought that those
dilaton-Maxwell-Einstein papers may be considered relevant to our
model in the asymptotic regions, where the scalar field settles to
its vacuum value, but even from this point of view there are
significant differences. The dilaton in the present system is a
metric component and not an additional field, so its coupling to
other fields is not arbitrary. Similarly, the ``electric'' or
``magnetic'' fields are also of gravitational origin, and since
they have specific source, the field equations are inconsistent
with the separate conditions $E_{r} = 0$ or $H=0$. Consequently,
the solutions presented here cannot be classified as either
electric or magnetic but always (for non-vanishing
self-interaction) have both kinds of fields or none of them. A
third difference is that most of the above-mentioned papers
introduce a non-vanishing cosmological constant, which we avoid
here.

\begin{figure}[!t]
   \begin{center}
      \includegraphics[width=13cm]{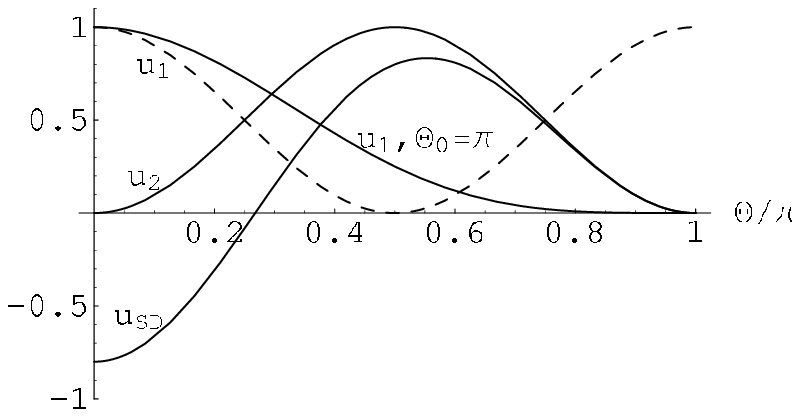} \\
   \caption{Dimensionless potential functions: a ``Higgs-like''
   $u_{1}(\Theta)$ with $\Theta_0=\pi/2$ -- dashed, the  ``single-well''
   $u_{1}(\Theta)$ with $\Theta_0=\pi$,
   the double well $u_{2}(\Theta)$ and the self-dual
   $u_{SD}(\Theta)$.}
 \label{fig4Pots}
     \end{center}
     \end{figure}

We solve the equations numerically taking the ``Kasner gauge''
$\alpha (r) e^{\sigma (r)}=1$ so that $g_{rr}=-1$. In order to get
localized solutions, we solve the equations with the boundary
conditions which ensure regularity at the origin and will be as
close as possible to asymptotically vacuum. That is, we impose on
the 3-metric
\begin{equation}
N(0)=1,\ \ \ N'(0)=0, \ \ \ \beta(0)=0 ,\ \ \ \beta'(0)=1, \ \ \
L_{\varphi}(0)=0,\ \ \ L_{\varphi}'(0)=0.
 \label{metricBC}\end{equation}
As for the gravito-electromagnetic components, $A_{0}(0)$ can be
any finite constant while
\begin{equation}
A_{0}'(0)=0 ,\ \ \ \
  \lim_{r\rightarrow\infty} A_{0}(r)= 0 , \ \ \ \ A_{\varphi}(0)=0
 \label{EMBC}\end{equation}
(for $A_{\varphi}(\infty)$ see below) and for the ``dilaton'' and
the scalar field
\begin{equation}
\sigma(0)=0,\ \ \ \sigma'(0)=0,\ \ \
\Theta(0)=0,\ \ \ \lim_{r\rightarrow\infty} \Theta(r)=\Theta_0
 \label{scalarBC}\end{equation}
where $\Theta_0$ is the location of the minimum of the potential.
The value of $\Theta_0$ distinguishes between two types of
hedgehog configurations: a ``sigma model hedgehog'' and a ``Higgs
hedgehog''. The first can be realized if the scalar tends
asymptotically to the south pole, i.e. $\Theta_0=\pi$. In this
case $A_{\varphi}(\infty)$ can be any finite constant and the
gravito-magnetic flux is not quantized. The second corresponds to
any other value $0<\Theta_0<\pi$ which is associated with a
quantized flux: $A_{\varphi}(\infty)=m/q$. In order to discuss
both kinds of solutions, we may use the following simple potential
function:
\begin{equation}
u_{1}(\Theta)=
\left(\frac{\sin^2(\Theta/2)-\sin^2(\Theta_0/2)}{\sin^2(\Theta_0/2)}\right)^2
 \label{SymmBrkPot}\end{equation}
where $\Theta_0$ is, as mentioned above, the location of the
potential minimum. It is actually the standard Higgs potential
multiplied by the conformal factor ${\cal E}(|\Phi|)$ and written
in terms of the angular field. The case $\Theta_0 =\pi$
corresponds to a ``single well potential''. We will also use the
``double well potential'', which is a special case of
(\ref{SDPotTheta}) with two minima at the two poles
$\Theta=0,\pi$:
\begin{equation}
u_{2}(\Theta)=\sin^2\Theta
 \label{DoubleWellPot}\end{equation}
These minima do not break the symmetry so they allow topological
solutions of the ``sigma model hedgehog'' type. This is the
potential used by Leese \cite{Leese1991} in $D=3$ flat spacetime
for obtaining the self-dual Q-lumps. Ward \cite{Ward2003} used
another form given by $u(\Theta)=\sin^2\Theta(1+\cos^2\Theta)$,
which we will not use here. All the other potentials are shown in
figure \ref{fig4Pots}.

\section{Special Solutions: Strings with Linear and/or Angular Momentum}
\setcounter{equation}{0} \label{SpecSol}

We solve numerically the field equations, using the MATHEMATICA
package, after rewriting them as a system of 14 coupled first order
ordinary differential equations. We use the dimensionless radial
variable $x=\mu r$, and rescale accordingly the other dimensional
quantities $q/\mu \rightarrow q$, $\omega/\mu \rightarrow \omega$,
 $\mu \beta (r) \rightarrow \beta (r)$ and
 $\mu L_{\varphi} (r) \rightarrow L_{\varphi} (r)$.

The following sections describe the main characteristics of the
solutions. In all cases we take $m=1$.

\subsection{Higgs-like Potentials}
\setcounter{equation}{0} \label{Symmetry breaking potentials}

\begin{figure}[!t]
   \begin{center}
      \includegraphics[width=13cm]{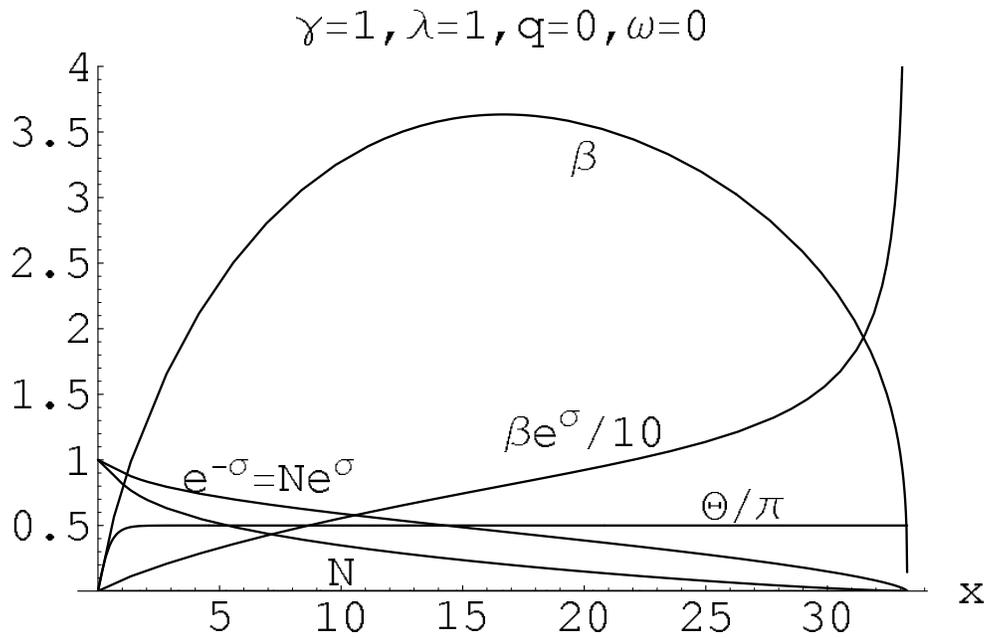} \\
   \caption{The fields of a static global string solution for
   a Higgs-like potential.}
 \label{figSSBom0KasG}
     \end{center}
     \end{figure}

\begin{figure}[!t]
   \begin{center}
      \includegraphics[width=13cm]{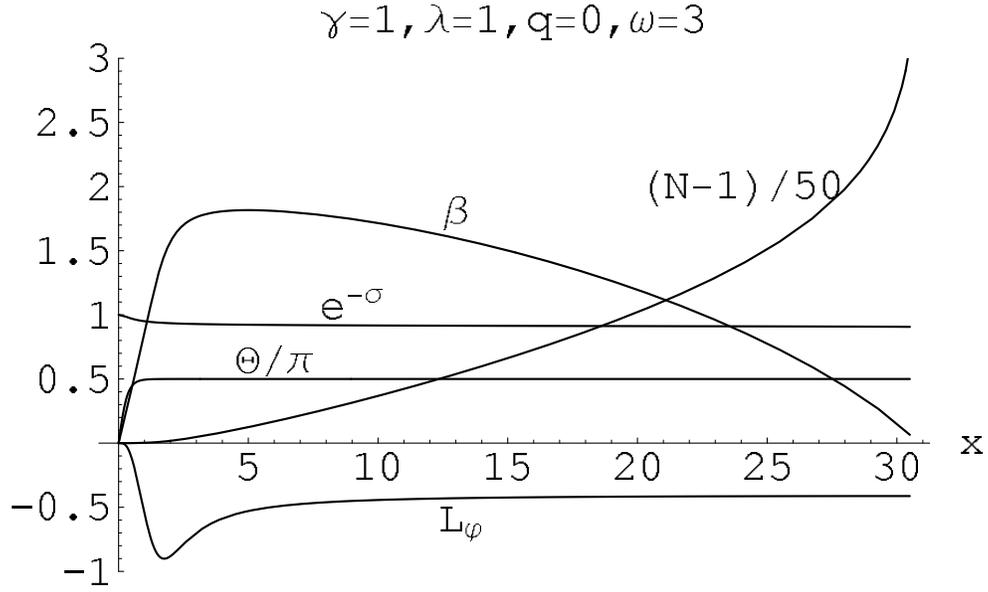} \\
   \caption{The fields of a purely rotating global string solution for
   a Higgs-like potential.}
 \label{figSSBom3KasG}
     \end{center}
     \end{figure}

We start by discussing a family of solutions for potentials of a
Higgs type, whose minima break spontaneously the $U(1)$ symmetry --
eq (\ref{SymmBrkPot}). The main disadvantage of the Higgs type
models is that the solutions are not localized ``enough'', so the
gravitational field is still singular, and we present them here
only as a basis for improvement. Figure \ref{figSSBom0KasG}
depicts the fields of a static solution for the case
$\Theta_{0}=\pi/2$. It is very similar to the Cohen-Kaplan
solution. For small $\gamma$ the singularity appears for very
large $r$ (or $x$ - for $\gamma = 0.2$ and $\omega=0$ the
singularity is already at $x \sim 10^{9}$), so for demonstrative
reasons we took quite a large $\gamma$. There are two kinds of
singularities. The first is the one appearing in figure
\ref{figSSBom0KasG}, where $N$, $e^{-\sigma}$ and $\beta$ vanish
while $\beta e^{\sigma}$ diverges for some finite $r$. For
$\omega=0$ $Ne^{\sigma}=e^{-\sigma}$ so $N$ and $e^{-\sigma}$
vanish at the same point. If $|\omega| > 0$ but smaller than a
certain bound (which depends on $\gamma$), $Ne^{\sigma}$ decreases
faster than $e^{-\sigma}$, so it is the zero of $Ne^{\sigma}$ or
rather $N$ which determines the singularity. If however $|\omega|$
is large enough, $\beta e^{\sigma}$ changes its trend and vanishes
while $N$ and $Ne^{\sigma}$ diverge. This last behavior is seen in
figure \ref{figSSBom3KasG}. A similar behavior is found for $q\neq
0$ as well.

This behavior is a direct result from the existence of the long
range Goldstone fields that exist in this model. This is the
origin for the singularity of the static metric of the global
string, and the other long range effects in other global defects.

As in the Cohen-Kaplan solution, it is the contribution to the
energy density from the angular dependence which does not vanish
fast enough to produce a localized source. In addition to the
angular dependence, we allow here also time dependence and
$z$-dependence which yield similar contributions all present in
the right-hand-side of (\ref{EinstN})-(\ref{EinsteinRotphiphi}).
Since all these terms are proportional to $\sin^2\Theta$, it is
quite easy to identify the way to improve the asymptotic behavior
namely, taking $\Theta_0 = \pi$ or $|\Phi|\rightarrow\infty$ as
$r\rightarrow\infty$. Note that the opposite direction,
$\Theta_0<<1$ gives the Higgs limit, as already mentioned.

\subsection{Single Well Potential}
In order to have solutions with $\Theta_0 = \pi$ (or
$|\Phi|\rightarrow\infty$), we take the following simple
monotonically decreasing potential
$U(|\Phi|)=\lambda\mu^4/(1+|\Phi|^2/\mu^2)$, which is one of the
``vacuumless'' potentials of Cho and Vilenkin
\cite{ChoVil1999a,ChoVil1999b,ChoVil2003}, who also found only
singular string-like solutions. However, since we use it within
the sigma model framework, this potential, unlike those of
 Cho and Vilenkin, is not vacuumless but a single-well potential
 with a single minimum at the south pole ($\Theta=\pi$) as is evident
 from its behavior in terms of the angular field:
\begin{equation}
u(\Theta)=\cos^2(\Theta/2)
 \label{SingleWellPot2}\end{equation}
Thus, we may get non-singular global strings and indeed we find
non-singular stationary solutions, though not static ones. There
are two kinds of string-like solutions: purely rotating (PR in
what follows) where no gravito-electromagnetic fields exist, and
rotating solutions with additional gravito-electromagnetic fields
(RGEM). On a first sight, one may think that the fields of the
latter may be either gravito-magnetic or gravito-electric,
according to the field components produced by the momentum along
the string axis. However, this is not the case (if
self-interaction exists) since the field equations are
inconsistent with either $E_{r} = 0$ or $H = 0$. Therefore, only
solutions with gravito-electromagnetic fields exist.

Both kinds of solutions (PR and RGEM) have $\omega\neq0$ and the
PR solutions have $q=0$. On the other hand, RGEM solutions exist
for either vanishing or non-vanishing values of $q$. The fact that
$q=0$ allows non-vanishing gravito-electromagnetic fields is a
direct result of the ``Maxwell'' equation for the ``displacement''
field $D^i$ eq (\ref{eqD}), which shows that it is $D^i$ (rather
than $E^i$) that vanishes if $q=0$. Since there is no
proportionality between these two gravito-electric fields, $D^i
=0$ does not imply $E^i = 0$ nor $B=0$ (or $H=0$) but only a
specific relation between them, somewhat similarly to a medium
with electric and magnetic permeabilities. Therefore, these fields
can be non-zero even when $q=0$. Another way to see that there
should exist RGEM solutions when $q=0$ is from the fact that it is
possible to ``gauge away'' $q$ by a coordinate transformation
which however will not transform to zero the
gravito-electromagnetic fields\footnote{we thank the referee for
pointing out towards this direction.}. The case $q=0$ splits
therefore into two branches: a PR branch and a RGEM branch.

\begin{figure}[!t]
   \begin{center}
      \includegraphics[width=13cm]{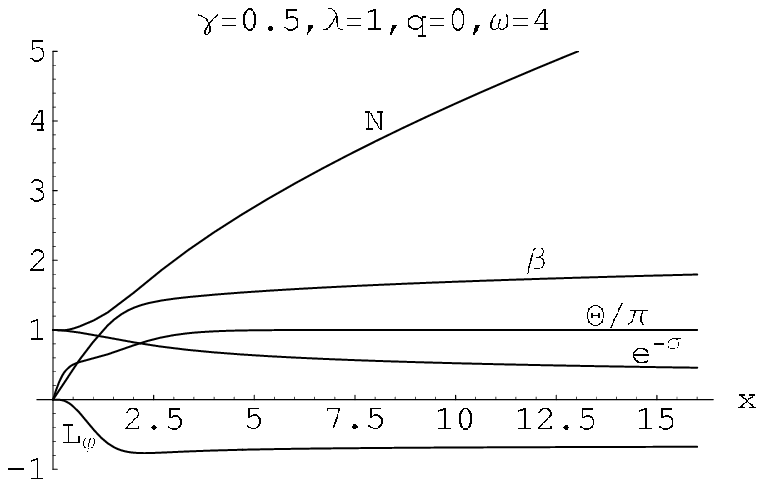} \\
   \caption{The fields of a purely rotating global string solution for a
   single well potential.}
 \label{fig1MinRot}
     \end{center}
     \end{figure}

\begin{figure}[!t]
   \begin{center}
 \includegraphics[width=13cm]{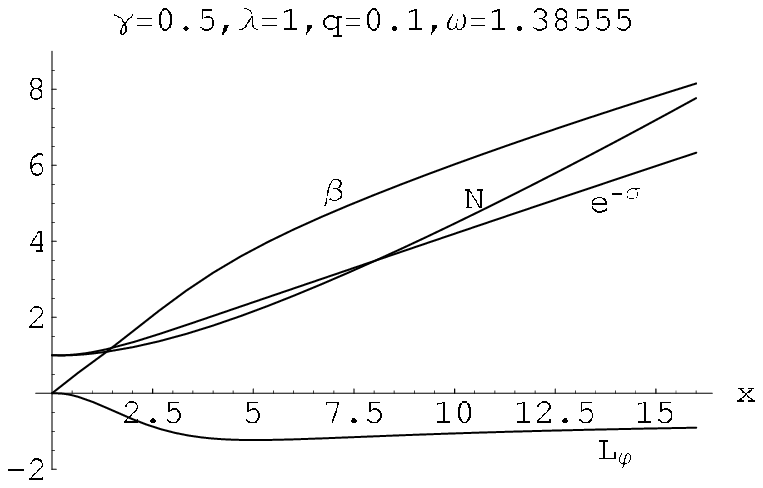} \\(a)\\
     \vspace{2cm}
      \includegraphics[width=13cm]{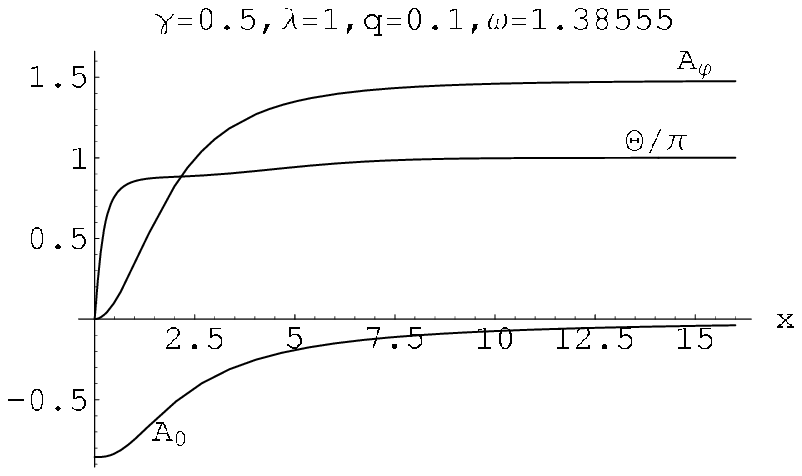}
      \\(b)\\
   \caption{The fields of a rotating gravito-electromagnetic global string solution for a
   single well potential.}
 \label{fig1MinRotEM2parts}
     \end{center}
     \end{figure}

The fields of a purely rotating solution are depicted in figure
\ref{fig1MinRot}. There is only a narrow range of $\omega$ values,
outside of which only singular solutions exist. Those of a
rotating gravito-electromagnetic solution are presented in the two
parts of figure \ref{fig1MinRotEM2parts}. RGEM solutions with
$q=0$ are similar and are not shown. The fact that RGEM
non-singular solutions exist with smaller $\omega$ is a general
one, and may be understood by the fact that less angular momentum
is required to balance the scalar and gravitational
self-attraction, in the presence of gravito-electromagnetic
fields. Another significant difference between the two kinds of
solutions, is the asymptotic behavior of $e^{-\sigma}$. While it
is monotonically decreasing function in the PR case, it is
monotonically increasing for RGEM solutions. This difference is a
direct consequence from the field equations: it is obvious from
equation (\ref{EinstSigm}) that for vanishing
gravito-electromagnetic fields, the function $\sigma(r)$ is
monotonically increasing so $e^{-\sigma(r)}$ is decreasing. On the
other hand, for RGEM solutions, the magnetic contribution is the
dominant term for large distances and it comes with an opposite
sign, so $e^{-\sigma(r)}$ has the opposite large distance behavior
- although it may start decreasing near the axis.

Another kind of potential which we can consider is
\begin{equation}
u(\Theta)=\cos^4(\Theta/2)
 \label{SingleWellPot4}\end{equation}
which is just the special case $\Theta_0=\pi$ of
(\ref{SymmBrkPot}). We find non-singular PR solutions, which are
quite similar to those of the previous single well potential, thus
we do not present them. However, we were able to obtain only
singular RGEM solutions for this potential.

\subsection{Double Well Potential}
\begin{figure}[!t]
   \begin{center}
      \includegraphics[width=13cm]{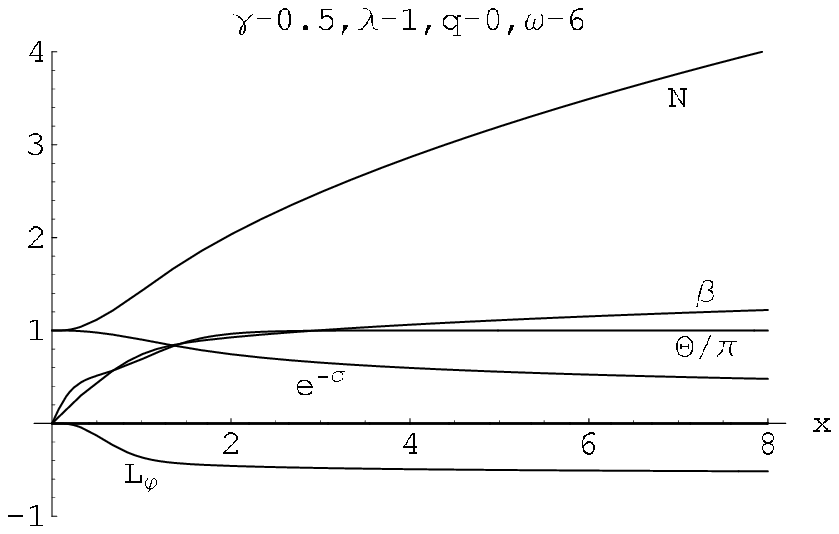} \\
   \caption{The fields of a purely rotating global string solution for a
   double well potential.}
 \label{fig2MinJ}
     \end{center}
     \end{figure}

\begin{figure}[!t]
   \begin{center}
    \includegraphics[width=13cm]{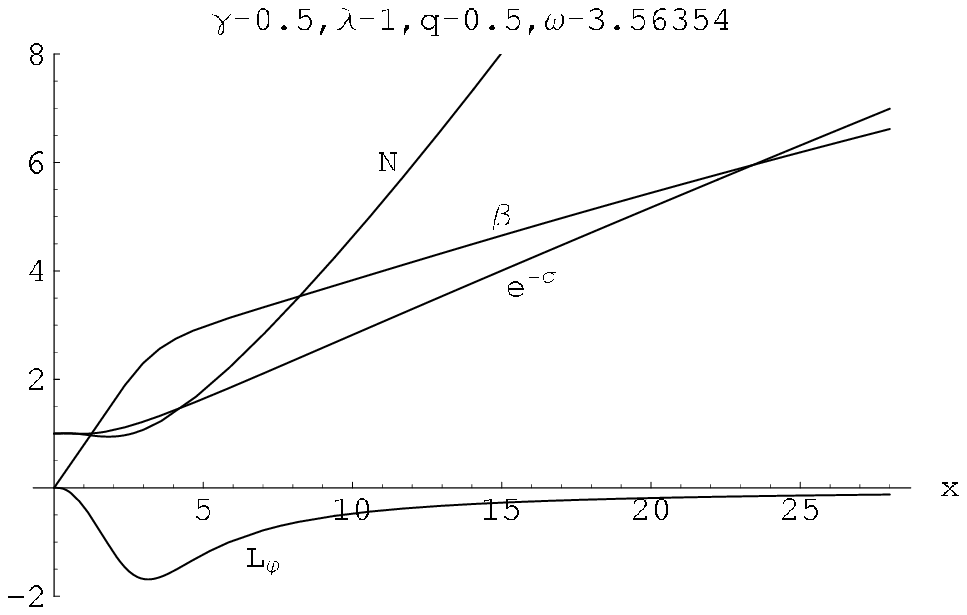}
    \\(a)\\
    \vspace{2cm}
      \includegraphics[width=13cm]{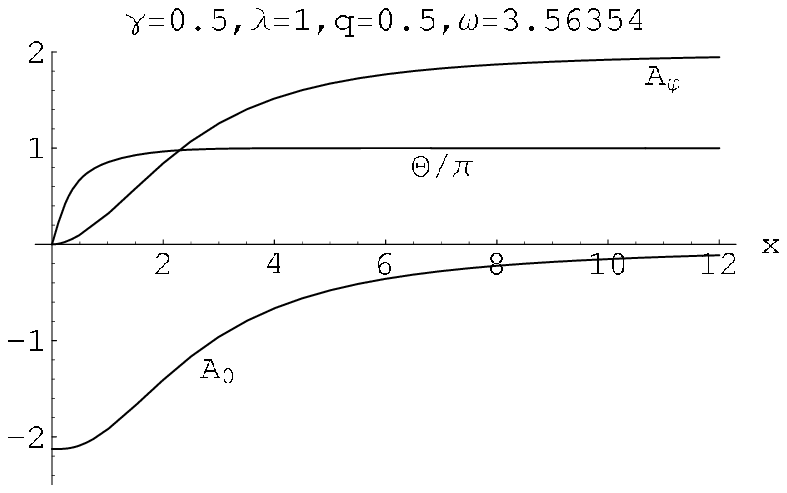}
      \\(b)\\
   \caption{The fields of a rotating gravito-electromagnetic global string solution for a
   double well potential.}
 \label{fig2MinEM2parts}
     \end{center}
     \end{figure}

We take the double well potential (\ref{DoubleWellPot}). The two
minima do not break the symmetry, so they allow topological
solutions of the ``sigma model hedgehog'' type. The fields of a
purely rotating solution appear in figure \ref{fig2MinJ}. As for
the previous potential, there is a range of $\omega$ values
outside of which only singular solutions exist,  while $\gamma$
can be quite large. The fields of a rotating
 gravito-electromagnetic global string are presented in figure
\ref{fig2MinEM2parts}. As before, the $q=0$ RGEM solutions are
similar as they are related to the $q\neq0$ ones by a coordinate
transformation. The gravito-electromagnetic solutions become
closed above a certain critical value of $\gamma$. The behavior of
$e^{-\sigma}$ is also analogous to that of the single well
potential: decreasing for the PR case and increasing for the RGEM
case.

\subsection{The Self-Dual Potential in $D=3,4$ }
\begin{figure}[!t]
   \begin{center}
      \includegraphics[width=13cm]{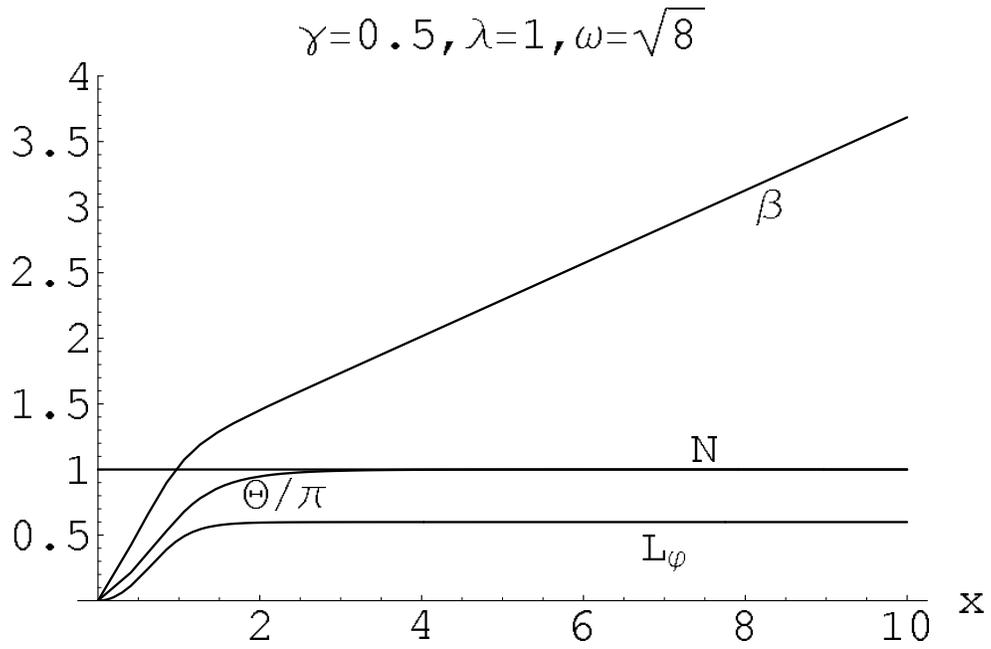} \\
   \caption{The fields of a rotating self-dual global string solution in $D=3$.
   Note that $N(x)=1$. }
 \label{figSD3Dsol}
     \end{center}
     \end{figure}

Now we take the potential (\ref{SDPotTheta}) which, as we saw
already, allows $D=3$ self-dual rotating strings or rather
vortices. We write it as in all other cases in terms of the
dimensionless potential function $u(\Theta)$, but with the $D=3$
coefficient\footnote{Note that $\kappa\mu^2=\gamma$ is still
dimensionless as in $D=4$ (or higher).} as $U=\lambda\mu^6
u(\Theta)$. The necessary condition for self-duality is therefore
$\omega^2=8\lambda\mu^4$. In order to obtain vortex solutions, we
may directly solve the second order system of equations
(\ref{EinstN}), (\ref{EinstL}), (\ref{EinsteinRotR}) and
(\ref{KGSpec}) with the substitutions $q=0$, $N=1$, $\sigma=0$,
$A_{0}=0$ and $A_{\varphi}=0$ which imply $E_r=0$ and $H=0$.
Alternatively, we may solve a first order system which include
(among other equations which we do not present), the equations
(\ref{SD3D}) and (\ref{SDell3D}) used for rotational symmetry. A
typical solution is shown in figure \ref{figSD3Dsol}.

Now we can proceed and search for solutions with the same
potential (up to powers of $\mu$) in $D=4$. It is not surprising
that we don't find self-dual solutions. Yet there exist
non-singular solutions, which we do not show since they are quite
similar to those of the double well potential.

\subsection{No Self-Interaction}
\begin{figure}[!t]
   \begin{center}
      \includegraphics[width=13cm]{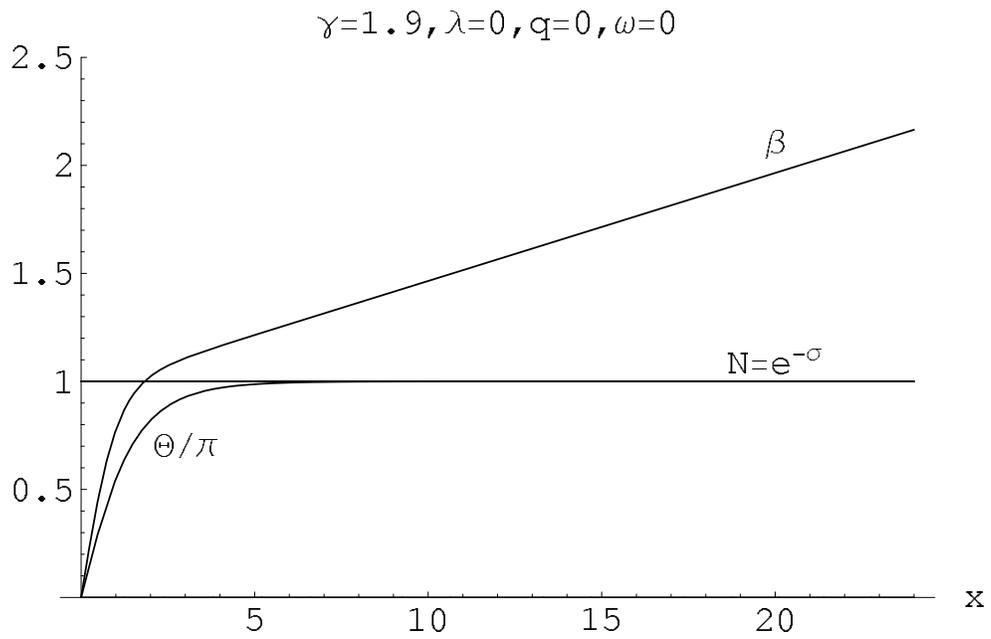} \\
   \caption{The fields of the Comtet-Gibbons (static) global string solution.}
 \label{figNoPotOmega0}
     \end{center}
     \end{figure}

For $U=0$ we reproduce the well-known Comtet-Gibbons string-like
solutions \cite{ComtetGibbons1988}, which have asymptotically a
conical geometry. Figure \ref{figNoPotOmega0} shows the fields of
a solution with a very large value of $\gamma$ that was taken in
order to stress the angular deficit, which is directly seen from
the asymptotic slope of $\beta(r)$.

\begin{figure}[!t]
   \begin{center}
      \includegraphics[width=13cm]{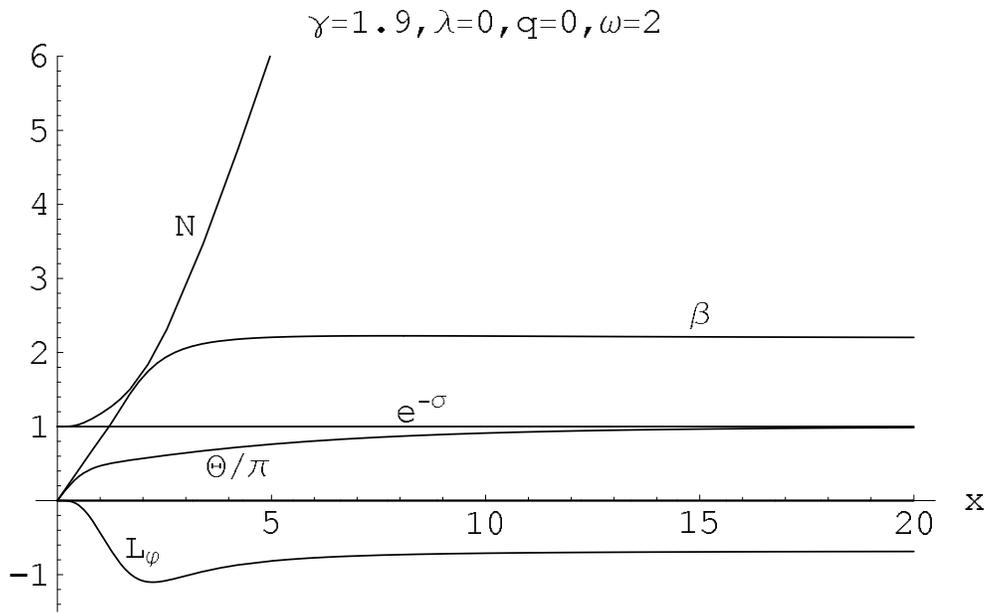} \\
   \caption{The fields of a purely rotating global string solution
   without self-interaction. Here $e^{-\sigma(x)}=1$.}
 \label{figNoPotOmegaNon0}
     \end{center}
     \end{figure}

A new type of (purely) rotating solutions is obtained for
non-vanishing $\omega$. The fields of one of these solutions are
depicted in figure \ref{figNoPotOmegaNon0}. All these solutions
have asymptotic cylindrical two-geometry, while the metric
component $N(r)$ increases linearly as $r\rightarrow\infty$.
Solutions exist for any $|\omega|>0$. The limit $\omega\rightarrow
0$ is singular, and does not reproduce the conical $\omega=0$
solutions.

\begin{figure}[!t]
   \begin{center}
     \includegraphics[width=13cm]{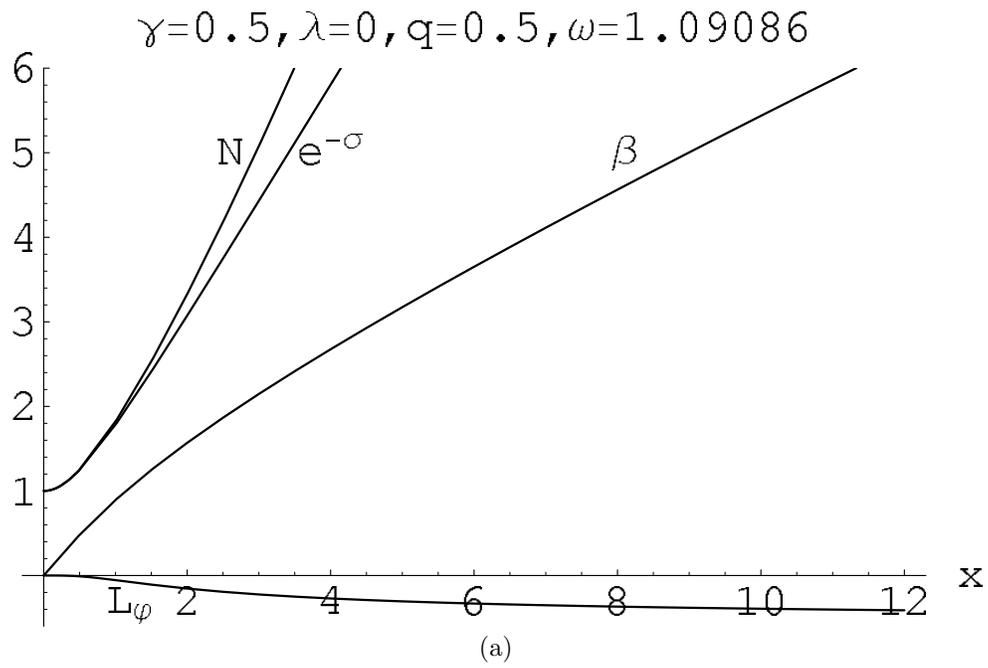} \\(a)\\
     \vspace{2cm}
      \includegraphics[width=13cm]{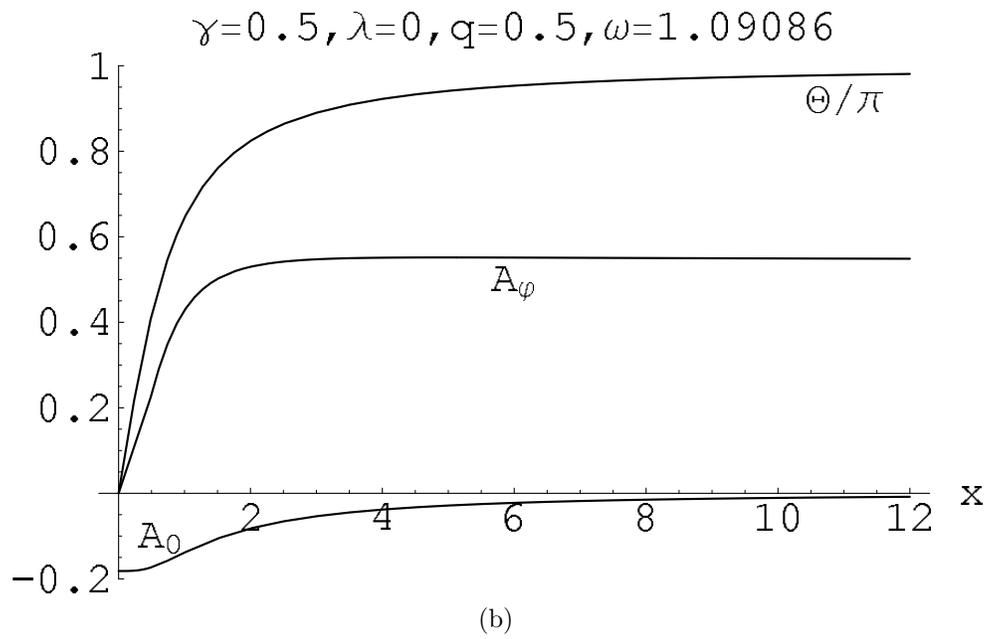} \\(b)\\
   \caption{The fields of a rotating gravito-electromagnetic global
   string solution without self-interaction.}
 \label{figNoPot2Parts}
     \end{center}
     \end{figure}

\begin{figure}[!t]
   \begin{center}
      \includegraphics[width=13cm]{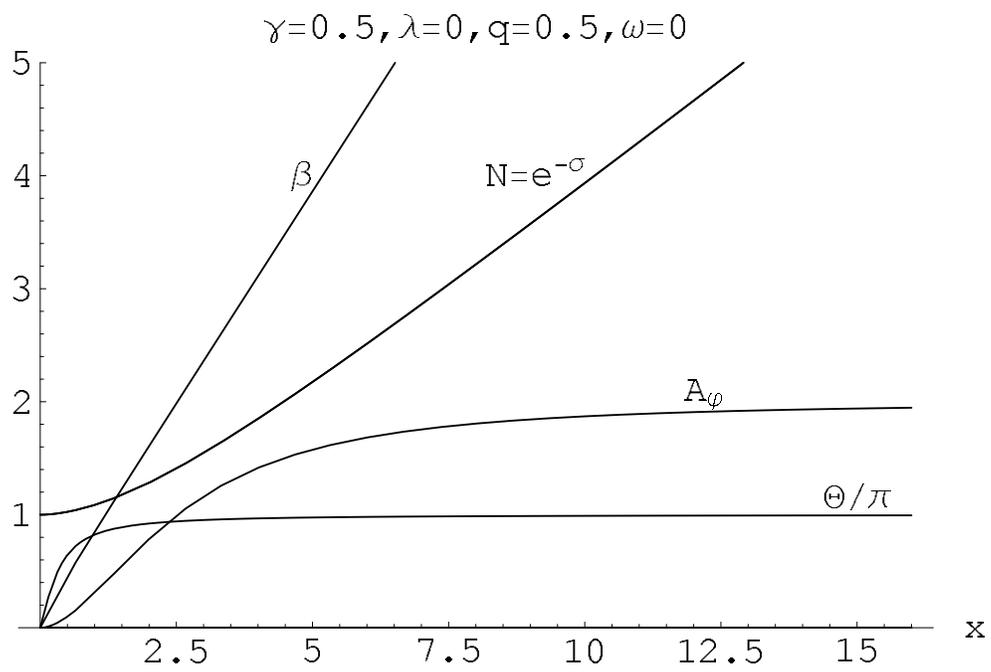} \\
   \caption{The fields of a non-rotating gravito-magnetic global string solution
   without self-interaction.}
 \label{figNoPotEM}
     \end{center}
     \end{figure}

The fields of a rotating gravito-electromagnetic global string are
presented in the two parts of figure \ref{figNoPot2Parts}. A
non-rotating solution with gravito-magnetic field is also found
and presented in figure \ref{figNoPotEM}. Because $\omega=0$ we
have here, as in the Comtet-Gibbons solution, $N=e^{-\sigma}$ but
now different from 1. This kind of purely gravito-magnetic
solution is special to the case of vanishing potential and cannot
exist otherwise.

\section{Conclusion and Outlook}\label{Conclusion}

We have shown that non-singular global cosmic string solutions
exist in a simple sigma model, if we also allow for rotation or
momentum current. Another result is the self-gravitating
generalization of the $D=3$ self-dual Q-lumps. Further
investigation is required in order to fully understand these
solutions and their consequences. The geometry of the solutions
should be clarified by geodesic analysis, and their possible role
in a cosmological context also deserves further study. Another
direction is a higher dimensional generalization of these
solutions to be codimension 2 branes. These may be relevant to the
extra dimensional resolution of the hierarchy problem, which was
advanced recently by several authors
\cite{CohenKaplan1999,Gregory1999,GiovanniniEtAl,GregorySantos2003}
using similar kind of solutions.

\pagebreak


\newpage

 \end{document}